\documentclass[final, aps, pra, floatfix, reprint, nofootinbib, citeautoscript, superscriptaddress, balancelastpage, longbibliography, showkeys]{revtex4-2}

\usepackage[T1]{fontenc}
\usepackage[english]{babel}
\usepackage{amsmath,amssymb,amsthm}
\usepackage{bm}
\usepackage{wasysym}
\usepackage{textcomp}
\usepackage[normalem]{ulem}
\usepackage{tabularx}
\usepackage[colorlinks, urlcolor=blue, citecolor=blue, linkcolor=blue, pdfstartview=FitH]{hyperref}
\usepackage[all]{hypcap}
\usepackage{graphicx}

\usepackage[dvipsnames]{xcolor}
\definecolor{MyColor}{rgb}{0.7, 0.0, 0.0}

\begin{document}

\title{Effective Conservation and Bistability of Atomic Alignment under Strong Spin~Exchange}

\author{A. K. Vershovskii}
\affiliation{Ioffe Institute, 194021 St.~Petersburg, Russia}


\begin{abstract}
We present a phenomenological model of anomalous alignment signals in dense cesium vapor under linearly polarized pumping and fast spin exchange near zero magnetic field. Despite the absence of a conservation law for rank-2 angular momentum, our recent experiments reveal anisotropic narrow resonances, hysteresis, and bistability. We attribute these effects to a stretched state forming a collective mode in which orientation and alignment are bidirectionally coupled. This mode acts as a reservoir, preserving the essential properties of alignment despite rapid spin exchange.
\end{abstract}
\maketitle

\section{Introduction}\label{sec:intro}

Optical pumping of alkali-metal vapors provides an example of how well-understood microscopic dynamics \cite{Omont_1977} can produce collective effects not anticipated from the single-atom picture, such as the spin-exchange relaxation-free (SERF) regime \cite{Happer_Tam_1977}.

SERF is a manifestation of the law of conservation of angular momentum: when the spin-exchange rate far exceeds the Larmor frequency, rapid collisions suppress transverse relaxation of the collective orientation, characterized by the relaxation time $T_2$, because spin exchange conserves total angular momentum \cite{Happer_Tam_1977, Appelt_Ben-AmarBaranga_Young_Happer_1999}. Under ordinary conditions, the longitudinal relaxation of the orientation, characterized by the relaxation time $T_1$, does not contain a spin-exchange contribution, as a consequence of the same conservation law.

This principle underpins ultrahigh-sensitivity zero-field magnetometry \cite{Kominis_Kornack_Allred_Romalis_2003, Ledbetter_Savukov_Acosta_Budker_Romalis_2008}. Using stretched states, i.e., states characterized by the maximum angular-momentum projection allowed for a given system \cite{Appelt_Ben-AmarBaranga_Young_Happer_1999}, partial suppression of spin-exchange relaxation has also been achieved at nonzero magnetic fields \cite{Budker_Romalis_2007, Scholtes_Schultze_IJsselsteijn_Woetzel_Meyer_2011,Schultze_Schillig_IJsselsteijn_Scholtes_Woetzel_Stolz_2017,Petrenko_Pazgalev_Vershovskii_2021}. The suppression of relaxation in a stretched state likewise follows from angular-momentum conservation.

Atomic alignment also offers practical advantages for magnetometry, including the absence of dead zones \cite{Ben-Kish_Romalis_2010, Wang_Wu_Xiao_Wang_Peng_Guo_2021}, reduced heading errors \cite{Hovde_Patton_Versolato_Corsini_Rochester_Budker_2011, Zhang_Kanta_Wickenbrock_Guo_Budker_2023}, and improved stability \cite{Rosner_Beck_Fierlinger_Filter_Klau_Kuchler_Rosner_Sturm_Wurm_Sun_2022}; its dynamics are described in \cite{Blum_2012, Omont_1977, Weis_Bison_Pazgalev_2006, Akbar_Kozbial_Elson_Meraki_Kolodynski_Jensen_2024, Meraki_Elson_Ho_Akbar_Kozbial_Kolodynski_Jensen_2023}. Alignment-based magnetometry has been explored near and away from zero magnetic field \cite{DiDomenico_Bison_Groeger_Knowles_Pazgalev_Rebetez_Saudan_Weis_2006, Breschi_Weis_2012, LeGal_Lieb_Beato_Jager_Gilles_Palacios-Laloy_2019, Meraki_Elson_Ho_Akbar_Kozbial_Kolodynski_Jensen_2023}.

Unlike \textit{orientation} (rank-1), \textit{alignment} (rank-2) is not associated with a nonzero mean angular momentum and therefore is not protected by conservation of total angular momentum; it generally relaxes through spin-exchange collisions \cite{Omont_1977, Happer_Tam_1977}. Our recent experiments \cite{Petrenko_Vershovskii_2025a, Petrenko_Vershovskii_2026a, Petrenko_Vershovskii_2026b}, however, challenged this expectation. In dense Cs vapor under linearly polarized pumping near zero field, we observed several anomalous features. At $T\lesssim 80^\circ$C, the signals approximately follow the standard theory \cite{Meraki_Elson_Ho_Akbar_Kozbial_Kolodynski_Jensen_2023}. As the temperature rises within the 80--100$^{\circ}$C range, anisotropy emerges: resonances scanned along one axis transverse to the beam remain on the scale expected for spin-exchange relaxation, whereas along the other transverse axis they exhibit SERF-like narrowing. With increasing temperature, a threshold-like behavior emerges near $T_T\approx 100^\circ$C: a memory-dependent effective field develops, leading to bistability with memory times reaching hundreds of seconds \cite{Petrenko_Vershovskii_2026a,Petrenko_Vershovskii_2026b}.

Nonlinear suppression of spin-exchange relaxation of higher-rank moments under strong circular pumping was previously demonstrated in \cite{katz_2013}. According to our current understanding, in this regime, the rank-2 component of a stretched state follows the slow dynamics of the orientation created by circular pumping. In our experiments and in the model developed to interpret them, we consider a distinct regime of linearly pumped Cs vapor in which one would ordinarily expect alignment without any orientation.

\begin{figure*}[!t]
\centering
\includegraphics[width=1.0\linewidth]{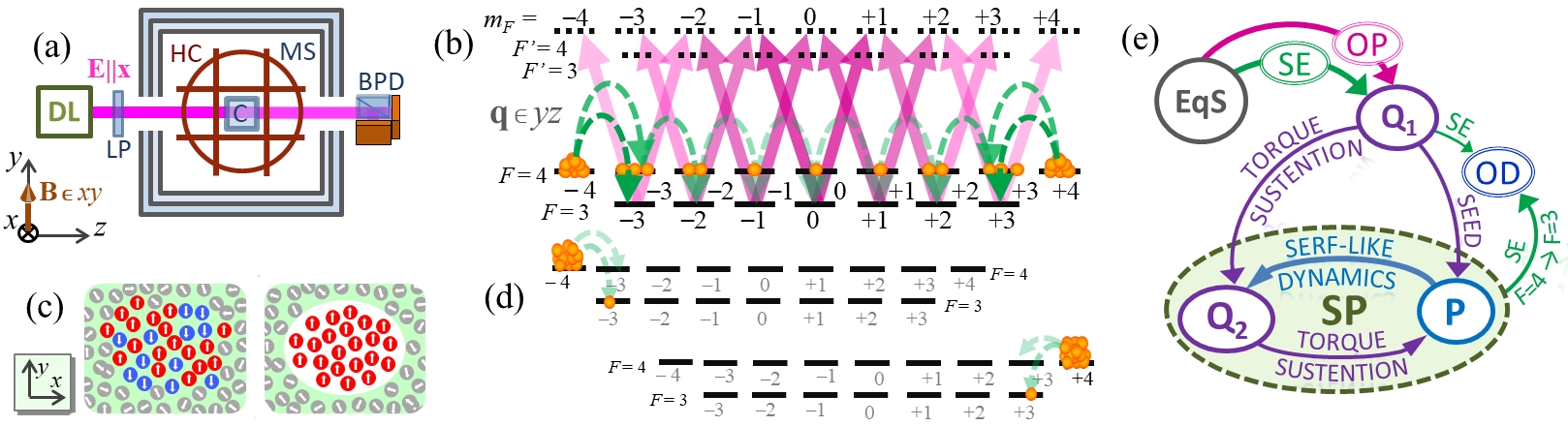}
\caption{(a) Typical experimental setup: DL, diode laser; LP, linear polarizer; MS, magnetic shield; HC, Helmholtz coils; C, cell; BPD, balanced photodetector. (b) Primary alignment $\mathsf{Q}_1$ in the $F=4$ manifold. Solid arrows indicate pumping, dashed arrows indicate spin-exchange. (c) Schematic formation of SP; the region of spin-exchange relaxation is shaded in green. (d) Two possible outcomes of the SP process: one of the two ``pockets'' (dark states) at $m_{Fy}=\pm F$ is populated. (e) Block diagram of the model: EqS, equilibrium state; OP, optical pumping; OD, optical detection; SE, spin-exchange; $\mathsf{Q}_1$, primary alignment; SP, spontaneous polarization; P, orientation component of SP; $\mathsf{Q}_2$, alignment component of SP (the secondary alignment).}
\label{fig:setup}
\end{figure*}

In \cite{Petrenko_Vershovskii_2026b} we introduced a minimal model that accounted for anisotropic narrowing by introducing spontaneous polarization (SP). This model also associated the observed slow relaxation of alignment with the dynamics of a stretched state, but it left several important features unexplained, most notably the symmetry breaking in the conversion of alignment into orientation, the effective field and the bistability of the system above the threshold. The threshold condition itself also remained unexplained. The model was intended primarily to demonstrate the basic mechanism, and its treatment of alignment dynamics was therefore restricted to the first components of the corresponding spherical tensor. Here we propose an extended phenomenological model that treats SP as a collective mode with bidirectional coupling between orientation and alignment [Fig.~\ref{fig:setup}(e)], providing a unified explanation for all significant anomalous effects observed in \cite{Petrenko_Vershovskii_2025a,Petrenko_Vershovskii_2026a,Petrenko_Vershovskii_2026b}.

The rest of the paper is organized as follows. Section~\ref{sec:system} summarizes the experimental regime. Section~\ref{sec:picture} introduces the basic principles of the proposed phenomenological model. Section~\ref{sec:model} formulates the proposed model mathematically. Section~\ref{sec:bistability} provides an explanation for bistability and hysteresis. Section~\ref{sec:results} presents the calculated results and compares them with the experimental observations. Section~\ref{sec:discussion} discusses the physical assumptions and limitations of the model. Section~\ref{sec:conclusion} summarizes the main conclusions. The Appendixes provide the tensor formalism and conventions and details of the mathematical derivation.

\section{Experimental regime and context}\label{sec:system}

The experimental setups used in \cite{Petrenko_Vershovskii_2025a,Petrenko_Vershovskii_2026a,Petrenko_Vershovskii_2026b} were typical for experiments of this kind and differed in details. In all three experiments, a cell containing Cs vapor and N$_2$ buffer gas was heated to a controlled temperature and placed in a very weak magnetic field, which was scanned around zero over a range of several tens to several hundred nanotesla, inside a magnetically shielded volume. The vapor was subjected to linearly polarized or very weakly elliptically polarized optical pumping. The transmitted pumping beam also served as the probe beam, allowing both absorption and polarization-rotation signals to be measured [Fig.~\ref{fig:setup}(a)]. The use of a single-beam geometry was motivated by the extreme sensitivity of the system to symmetry-breaking perturbations introduced by additional beams or by additional polarization components. This sensitivity was demonstrated, in particular, in \cite{Petrenko_Vershovskii_2026a}, where a small circular-polarization component, corresponding to an ellipticity of a few tenths of a degree, intentionally added to the linearly polarized pump, caused a sharp change in the observed signals.

Throughout the description of these experiments, we use a coordinate system with $\mathbf E\parallel\mathbf x$ and $\mathbf k\parallel\mathbf z$. This choice accommodates the possibility of circular pumping in addition to linear pumping \cite{Petrenko_Vershovskii_2026a} and therefore differs, for example, from the convention used in \cite{Meraki_Elson_Ho_Akbar_Kozbial_Kolodynski_Jensen_2023}.

The experiment \cite{Petrenko_Vershovskii_2025a}, which first reported anomalous suppression of alignment relaxation under linearly polarized pumping, covered a wider range of magnetic fields, up to 300~nT, and demonstrated a gradual narrowing of the alignment resonances as the magnetic field was reduced. The dependence of the alignment relaxation rate on magnetic field was anisotropic: under scan along the direction parallel to the $x$-axis, it showed a dependence on magnetic field resembling that of SERF resonances \cite{Happer_Tam_1977}.

In the most relevant and detailed experiment, performed with linear pumping \cite{Petrenko_Vershovskii_2026b}, a $5\times5\times5$~mm$^3$ glass cell containing Cs and 200~Torr of nitrogen (N$_2$) was placed at the center of 3D Helmholtz coils in a multilayer magnetic shield. 
The cell temperature $T$ was selected within the range of $77$--$103^\circ$C, where alignment signals with a sufficient signal-to-noise ratio were observed.
In this range, the atomic density was $(0.33$--$1.77)\times10^{13}$~cm$^{-3}$ and the spin-exchange rate was $R_{\mathrm{SE}}=2300$--$12900$~s$^{-1}$ \cite{Happer_Tam_1977, Seltzer_2008}.

The Cs vapor was subjected to linearly polarized pumping, resonant with the $F=I-1/2=3\leftrightarrow F'=I+1/2=4$ transition. The input light power was 3~mW with a beam cross section of 0.1~cm$^2$, corresponding to a pumping rate of $R_P\approx3\times10^{4}~s^{-1}$; it depleted the $F=3$ manifold and created an alignment on $F=4$ [Fig.~\ref{fig:setup}(b)].

A Glan prism was used to purify the pump polarization. Residual circular polarization, introduced by the optical elements, was compensated with an accuracy better than 0.2$^{\circ}$ using a quarter-wave plate. Ellipticity was monitored with a polarimeter; the final criterion for linearity was the absence of effects introduced by circular pumping \cite{Petrenko_Vershovskii_2026a}. The quarter-wave plate is not shown in Fig.~\ref{fig:setup}(a) to avoid suggesting that circularly polarized pumping was intentionally used.

A magnetic field $\mathbf B$ was applied in the $xy$-plane, with $B=|\mathbf B|\leq50$~nT. In the near-zero-field region, $\gamma B\ll R_{\mathrm{SE}}\ll R_P$, where $\gamma$ is the gyromagnetic ratio. 
Under these conditions, the SERF regime for \textit{orientation} is realized \cite{Happer_Tam_1977, Kominis_Kornack_Allred_Romalis_2003}. Experiments \cite{Petrenko_Vershovskii_2025a,Petrenko_Vershovskii_2026a,Petrenko_Vershovskii_2026b}, however, revealed narrow \textit{alignment} resonances under scan along along the $x$ axis with halfwidths as small as $3.3$~nT at $T=93^\circ$C, corresponding to an effective relaxation rate of $\approx70$~s$^{-1}$. Taking into account the decrease in the gyromagnetic ratio in zero field \cite{Happer_Tam_1977, Seltzer_2008}, this estimate decreases to $\approx30$~s$^{-1}$, corresponding to a relaxation time of $\approx33$~ms, which is of the same order as the
relaxation times of $\approx20$~ms measured in
\cite{Petrenko_Vershovskii_2025a}.

\section{Physical picture and notation}\label{sec:picture}

The physical picture underlying our model is based on two main components of the state created by optical pumping: the primary alignment directly produced by the pump and the spontaneous polarization (SP) \cite{Fortson_Heckel_1987,Klipstein_Lamoreaux_Fortson_1996,Andalkar_Warrington_Romalis_Lamoreaux_Heckel_Fortson_2002}. The latter can be decomposed into distinct spherical moments \cite{Omont_1977,Happer_Tam_1977}. We retain two moments that directly interact with the light: orientation (rank 1, dipole moment) and alignment (rank 2, quadrupole moment), which we refer to below as the secondary alignment. 
The signal contains a rapidly relaxing component (rate $\Gamma_A$) associated with the primary alignment and a slow component (rate $\Gamma_P$) associated with the collective mode.

We use angular-frequency variables $\boldsymbol{\omega}\equiv\gamma\mathbf B$ and $\omega_i=\gamma B_i$, where $i=x,y,z$.
As shown in \cite{Petrenko_Vershovskii_2025a, Petrenko_Vershovskii_2026a}, the presence of a field along the $z$ axis does not reveal significantly new effects; it merely leads to broadening and a reduction in amplitude of the narrow resonances observed at $B_z=0$. Further we restrict ourselves to considering fields in the $xy$ plane. 

The orientation, which is the leading component of the spontaneous polarization, is denoted by $\mathbf P=P\mathbf n_P$, with $0\leq P\leq1$. For $P>0$, the unit vector $\mathbf n_P\equiv\mathbf P/P$ specifies its direction. The primary and secondary alignments are represented by symmetric traceless Cartesian rank-2 tensors $\mathsf Q_1$ and $\mathsf Q_2$, respectively. For a uniaxial alignment, we denote its signed amplitude in the representation $\mathsf Q_j \equiv A_j(\mathbf n_j\mathbf n_j-\mathsf I/3)$ by $A_j$. We denote the projection of the alignment tensor $\mathsf Q_j$ onto an arbitrary direction $\mathbf m$ by $Q_{j,mm}\equiv\mathbf m\cdot\mathsf Q_j\cdot\mathbf m$, where $\mathbf m$ is a unit vector. We denote the equilibrium value of $\mathsf Q_1$ produced by the light by $\mathsf Q_{10} \equiv A_{10}\left(\mathbf e_x\mathbf e_x-\frac13\mathsf I\right)$.

The following qualitative considerations summarize the main assumptions underlying the model. While some of them follow directly from the pumping geometry and angular-momentum structure, others constitute hypotheses that are tested by comparison with the experimental observations. The latter include the following assumptions: that the orientation is replenished by the primary alignment through coupling with the secondary alignment; that the magnitude of the orientation remains approximately constant within the angular range in which the collective mode exists; and that the effective magnetic field can be represented by the torque exerted on the orientation by the primary alignment through the same coupling with secondary alignment. Explanations of line narrowing and bistability are, in turn, consequences of these assumptions.

\begin{figure}[!t]
\centering
\includegraphics[width=0.7\linewidth]{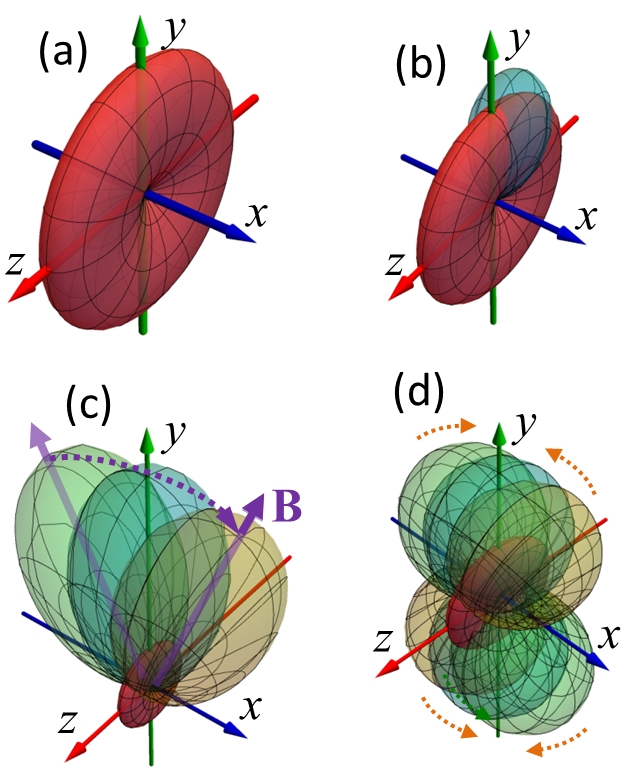}
\caption{Illustrative visualization of the angular-momentum distribution: (a) Primary alignment $\mathsf Q_1$. (b) Growth of orientation in the $yz$ plane from a small random seed. (c) Fully stretched state: rotation with $\mathbf{B}$ in the $xy$ plane. (d) The quadrupole component $\mathsf Q_2$ of the stretched state: rotation with $\mathbf{B}$ in the $xy$ plane. The orange arrows show the torque from $\mathsf Q_1$. The visualization was performed using the ADM software package \cite{ADM}.}
\label{fig:sp_process}
\end{figure}

The primary alignment $\mathsf Q_1$ in the $F=4$ manifold is negative ($A_1<0$). Upon quantization along the $x$-axis, projection of $\mathsf Q_1$ is negative, whereas along the $y$- and $z$-axes, it is positive, so that $Q_{10,xx}<0$ and $Q_{10,yy}=Q_{10,zz}>0$ [Figs.~\ref{fig:setup}(b) and \ref{fig:sp_process}(a)]. A small random polarization seed in the $yz$ plane may initiate formation of a stretched state [Fig.~\ref{fig:sp_process}(b)], which we identify with the spontaneously polarized collective mode \cite{Fortson_Heckel_1987,Klipstein_Lamoreaux_Fortson_1996,Andalkar_Warrington_Romalis_Lamoreaux_Heckel_Fortson_2002}. In particular, along $y$ the population accumulates in two dark ``pockets'' at $m_{Fy}=\pm F$ [Fig.~\ref{fig:setup}(b)]. As long as both pockets are equally populated, spin-exchange collisions efficiently deplete them in pairs and, after one or more collisions, redistribute atoms toward the $F=3$ manifold, from which they reenter the optical-pumping cycle.

Once any imbalance appears, atoms in the more populated state begin to lack collision partners, while pumping continues to replenish this state population. Under certain conditions, such feedback leads to formation of one or more SP regions with $\langle m_{Fy}\rangle\approx\pm F$ [Figs.~\ref{fig:setup}(c), \ref{fig:setup}(d), \ref{fig:sp_process}(b), and \ref{fig:sp_process}(c)]. Weak overlap of the optical profiles of the hyperfine levels and partial conservation of the nuclear spin in the $6^2P_{1/2}$ state are factors contributing to the pumping efficiency. 

The angular-momentum balance can be accommodated by N$_2$ molecules \cite{Klipstein_Lamoreaux_Fortson_1996}, which carry rotational angular momentum. This process differs from alignment-to-orientation conversion \cite{Rochester_Ledbetter_Zigdon_Wilson-Gordon_Budker_2012} in that the latter is a reversible transformation between atomic polarization moments, whereas the present mechanism involves an effectively irreversible pumping cycle.

For an ideal stretched state, the principal axis of the secondary alignment $\mathsf Q_2$ is parallel to $\mathbf P$, and  its amplitude is related to the polarization by $A_2=\sigma P>0$, where $\sigma$ is a constant (Appendix~\ref{app:stretched_moments}). This equality does not necessarily hold for an arbitrary nonequilibrium collective state. In general, the dynamics of $\mathsf Q_2$ can differ from that of $\mathbf P$. 
One significant reason for this is the relaxation of SP: if all atoms in the given volume were involved in the SP, no population capable of sustaining independent primary alignment would remain outside the collective mode.
The finite relaxation of SP therefore allows the collective and primary-alignment components to coexist and provides a natural channel for their continuous replenishment. In what follows, we use the stretched-state relation as an approximation for the slow collective mode, which provides grounds for associating its characteristic relaxation rate with the orientational rate $\Gamma_P$.

\begin{figure}[!t]
\centering
\includegraphics[width=0.75\linewidth]{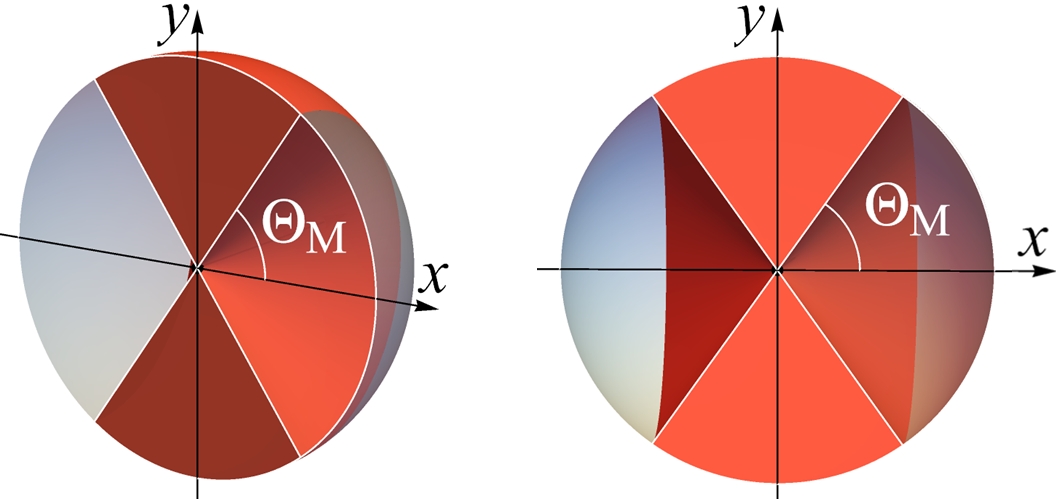}
\caption{Angular distribution of possible directions of spontaneous polarization (shaded in red), according to Eq.~\eqref{eq:magic}.}
\label{fig:angular}
\end{figure}

The projection of the primary alignment onto the secondary one is positive as long as the principal axis of $\mathsf Q_2$, which in the saturated stationary state is parallel to $\mathbf P$, lies within the angular region in which $Q_{1,n_Pn_P}$ is positive. 
For $\mathsf Q_1=\mathsf Q_{10}$, this region is a ``belt'' centered in the $yz$-plane:
\begin{equation}
\theta\in(\theta_M,\pi-\theta_M),
\quad
\theta_M=\arccos(1/\sqrt3).
\label{eq:magic}
\end{equation}
$\theta_M\approx54.74^\circ$
is the so-called magic angle measured from the $x$ axis \cite{Budker_Kimball_DeMille_2004}.
Consequently, replenishment of the secondary alignment, and hence of SP, can occur only within this belt, the intersection of which with the $xy$ plane consists of two equal, oppositely directed angular intervals [Fig.~\ref{fig:angular}].

The size of an SP region [Fig.~\ref{fig:setup}(c)] varies depending on the system parameters, primarily the beam intensity and profile. Weak residual spin-exchange relaxation partially transfers the population from the $F=4$ to the $F=3$ manifold, where it is detected optically; the loss of oriented atoms in $F=4$ is compensated by pumping through the primary alignment. The magnitude of $\mathbf P$ is determined by the balance between relaxation and nonlinear replenishment, and is ultimately limited by a finite number of atoms in the pumped volume. At a fixed pumping rate, we therefore treat $P$ as approximately constant within the angular range in which the collective mode is established. The SERF-protected orientation stabilizes $\mathsf Q_2$ against spin-exchange-induced destruction and determines its dynamics [Fig.~\ref{fig:setup}(e)].

An important feature of the present system is that the direction of $\mathbf P$ is not imposed by the optical pumping. This mechanism differs from conventional SERF schemes, in which the steady-state orientation is determined by the pumping direction and its deflection by the magnetic field. Therefore, in the quasi-stationary approximation, once established, the orientation $\mathbf P$ adiabatically follows the magnetic field $\mathbf B$, remaining practically parallel $(\uparrow\uparrow)$ or antiparallel $(\uparrow\downarrow)$ to it  \cite{Fortson_Heckel_1987}.

Transverse relaxation damps the precessional deviations of $\mathbf P$ from the field direction, while self-organization replenishes the longitudinal component. This separation is represented explicitly in the collective-mode equation below. For slow field scans, the angular velocity $\omega_B$ of the magnetic-field direction is assumed to be small compared with the transverse relaxation rate, $\omega_B\ll\Gamma_P$. Away from zero field, precessional adiabatic following additionally requires $\omega_B\ll\omega$, where $\omega\equiv|\boldsymbol{\omega}|$.Near a vanishing total field, its direction is no longer defined and the following approximation can fail. This may modify the local trajectory of $\mathbf P$ and cause a transient deviation from the $xy$ plane.

The Zeeman energy difference between the two stretched states $(\uparrow\uparrow)$ and $(\uparrow\downarrow)$ is $\hbar(2I+1)\gamma B$ and is many orders of magnitude smaller than the thermal energy $k_BT$. For example, at $B=50$~nT and $T=373$~K, their ratio is approximately $1.8\times10^{-10}$ for $\gamma/(2\pi)=3.5$~Hz/nT. Consequently, either direction in the angular range given by Eq.~\eqref{eq:magic} can be selected by an initial fluctuation.

Another difference from classical SERF schemes concerns the optical signal. In conventional magnetometry, orientation is detected either through circular dichroism (CD) in the absorption of circularly polarized probe light or through circular birefringence (CB), which rotates the polarization plane of linearly polarized probe light detuned from resonance. In our case, the dominant observable is linear dichroism (LD). This is demonstrated by the strong, narrow absorption resonances observed with linearly polarized light \cite{Petrenko_Vershovskii_2026b}, which cannot be accounted for by CB or CD alone \cite{Meraki_Elson_Ho_Akbar_Kozbial_Kolodynski_Jensen_2023}. The reduced relaxation of $\mathsf Q_2$ allows it to dominate the signal at high atomic densities.

A distinctive feature at temperatures exceeding the threshold value is the angular boundary of the LD signal. We attribute this to the condition Eq.~\eqref{eq:magic}. 
The threshold dependence and nonlinearity of this effect suggest that, above threshold, exiting the angular zone boundaries not merely damps replenishment of SP but instead completely suppresses SP by switching the sign of the feedback responsible for sustaining it.

The coupling of $\mathbf P$ to the primary alignment $\mathsf Q_1$  is mediated by the secondary alignment $\mathsf Q_2$. In a saturated state, when the magnitude of $\mathbf P$ is nearly constant, this coupling can be represented as a state-dependent precession torque acting on $\mathbf P$. 
At small $B$ and sufficiently high atomic density,  this torque, together with relaxation, shifts the stationary direction of $\mathbf P$ toward the $y$ axis.
This contribution is represented by $B_{y,\mathrm{eff}}$, so that the total effective field is $\mathbf B'=\mathbf B+B_{y,\mathrm{eff}}\mathbf e_y$.
In the quasi-stationary approximation, $\mathbf P$ follows $\mathbf B'$.

For a scan in the $xy$ plane with a finite transverse component $B_x$, $\mathbf P$ can follow a quasistationary branch parallel
or antiparallel to $\mathbf B'$ only while the replenishment condition in Eq.~\eqref{eq:magic} remains satisfied. In this geometry, the condition becomes $|B_y'|>\sqrt{2}|B_x|$
(see Appendix~\ref{app:angular_threshold}). Consequently, a continuously followed polarized branch cannot pass through $B_y'=0$ while remaining within the
allowed angular region.

Within the assumed angular-restriction model, at $|B_y'|<\sqrt{2}|B_x|$ the spontaneous polarization is suppressed, and memory carried by the collective mode is lost. An effective field shifts the ranges of the applied field $\mathbf B$ in which the polarized branches exist, towards zero $|B_y|$ values without changing the magic-angle zone [Fig.~\ref{fig:angular}] itself. For sufficiently strong $B_{y,\mathrm{eff}}$, these ranges overlap, allowing preservation of SP near $B_y=0$ and thus providing a geometric basis for bistability and history-dependent switching.

Thus, the sharp threshold for memory effects in this model is related to the balance between the magnitude of the effective field and the magic-angle condition.

\section{Phenomenological model
}\label{sec:model}

\subsection{Primary alignment}

The primary alignment produced directly by the optical pumping is represented by the symmetric traceless tensor $\mathsf Q_1$. In the absence of a magnetic field, the linearly polarized pump establishes a stationary alignment $\mathsf Q_{10}$ with its symmetry axis along the light polarization, $\mathbf e_x$. The magnetic evolution of atomic alignment and its relaxation toward the pump-prepared state have been treated previously within the irreducible-tensor formalism~\cite{Breschi_Weis_2012}. Following this approach, we describe the primary alignment by
\begin{equation}
\dot{\mathsf{Q}_1}
=
\Omega\mathsf{Q}_1
-
\mathsf{Q}_1\Omega-
\Gamma_A
\left(
\mathsf{Q}_1-\mathsf{Q}_{10}
\right),
\label{eq:Q1_main}
\end{equation}
where $\Omega$ is the antisymmetric rotation operator defined by $\Omega\mathbf v=\boldsymbol{\omega}\times\mathbf v$ for an arbitrary vector $\mathbf v$. The tensor rotation and the stationary solution are given explicitly in Appendixes~\ref{app:tensor} and~\ref{app:primary}, respectively.

The components relevant to the collective response are
\begin{align}
Q_{1,xy}
&=
A_{10}
\frac{
\omega_x\omega_y
\left(
\Gamma_A^2+4\omega_x^2-2\omega_y^2
\right)
}{
\left(
\Gamma_A^2+\omega_x^2+\omega_y^2
\right)
\left(
\Gamma_A^2+4\omega_x^2+4\omega_y^2
\right)
},
\label{eq:Q1xy_main}
\\
Q_{1,xz}
&=
-
A_{10}
\frac{
\Gamma_A\omega_y
\left(
\Gamma_A^2+4\omega_x^2+\omega_y^2
\right)
}{
\left(
\Gamma_A^2+\omega_x^2+\omega_y^2
\right)
\left(
\Gamma_A^2+4\omega_x^2+4\omega_y^2
\right)
}.
\label{eq:Q1xz_main}
\end{align}

The stationary solutions Eqs.~\eqref{eq:Q1xy_main} and \eqref{eq:Q1xz_main} coincide with the corresponding results of Meraki et al.~\cite{Meraki_Elson_Ho_Akbar_Kozbial_Kolodynski_Jensen_2023} for the case $B_z=0$.

\subsection{Collective polarization}

The collective polarization is described phenomenologically by
\begin{equation}
\dot{\mathbf P}
=
\boldsymbol{\omega}\times\mathbf P
-\Gamma_P\mathbf P
+\boldsymbol{\alpha}(\mathbf P).
\label{eq:P_general}
\end{equation}
To preserve the axial symmetry characteristic of Bloch-type dynamics with respect to the precession vector, this term can be written in the form
\begin{equation}
\boldsymbol{\alpha}(\mathbf P)
=
\chi(P)\,
\mathsf\Pi_{\parallel}\mathbf P,
\quad
\mathsf\Pi_{\parallel}
=
\mathbf n_{\omega}\mathbf n_{\omega}^{\mathsf T},
\quad
\mathbf n_{\omega}
=
\frac{\boldsymbol{\omega}}{|\boldsymbol{\omega}|},
\label{eq:alpha}
\end{equation}
where $\chi(P)$ is the self-organization gain. We assume that this gain replenishes the polarization component parallel to the precession vector, whereas the transverse components remain subject to relaxation. The projector is defined for $|\boldsymbol{\omega}|>0$. Thus, the collective-polarization equation becomes
\begin{equation}
\dot{\mathbf P}
=
\boldsymbol{\omega}\times\mathbf P
-\Gamma_P\mathbf P
+\chi(P)\,
\mathsf\Pi_{\parallel}\mathbf P .
\label{eq:P_main}
\end{equation}

The coupling between the uniaxial tensors $\mathsf Q_1$ and $\mathsf Q_2$ can be characterized by their scalar overlap (the Frobenius product), which provides a rotationally invariant measure of their relative alignment:
\begin{equation}
\Delta Q\equiv\mathsf Q_1:\mathsf Q_2
=\operatorname{Tr}(\mathsf Q_1^T\mathsf Q_2)
=
-\frac{2}{3}|A_1A_2| L_2(\cos\theta),
\label{eq:overlap}
\end{equation}
where $\theta$ is the angle between their principal axes, and $L_2(x)$ is the second Legendre polynomial.

\begin{equation}
L_2(x)=\frac{1}{2}(3x^2-1).
\label{eq:legendre}
\end{equation}

In accordance with this overlap picture, one should expect that, away from saturation, the self-organization gain should contain the same angular factor,
\begin{equation}
\chi \propto -L_2(\cos\theta)
=-\frac{1}{2}(3\cos^2\theta-1).
\label{eq:chi_overlap}
\end{equation}
The actual dependence of the replenishment on the polarization magnitude near saturation is unknown, however, and this expression therefore cannot be used in the subsequent calculations. It nevertheless provides a qualitative basis for the angular dependence of the self-organization effect and motivates the appearance of the second-rank angular factor $L_2(\cos\theta)$ in the model.

The self-organization of atomic moments and the emergence of states with the polarization parallel or antiparallel to the magnetic field have been considered previously~\cite{Fortson_Heckel_1987, Klipstein_Lamoreaux_Fortson_1996,Andalkar_Warrington_Romalis_Lamoreaux_Heckel_Fortson_2002}. In particular, an explicit solution for the one-dimensional case was obtained in Ref.~\cite{Fortson_Heckel_1987}. A corresponding three-dimensional treatment is beyond the scope of the present phenomenological model.

Eq.~\eqref{eq:P_main} can be written as
\begin{equation}
\begin{aligned}
\dot{\mathbf P}
=\boldsymbol{\omega}\times\mathbf P-\Gamma_P\mathsf\Pi_\perp\mathbf P
+[\chi(P)-\Gamma_P]\mathsf\Pi_\parallel\mathbf P,
\end{aligned}
\end{equation}
where $\mathsf\Pi_\perp=\mathsf I-\mathsf\Pi_\parallel$. For a nonzero stationary solution,
\begin{equation}
\begin{aligned}
P&=P_0,
\quad
\boldsymbol{\omega}\times\mathbf n_P=0,
\quad
\chi(P_0)=\Gamma_P.
\end{aligned}
\label{eq:P_main_sat}
\end{equation}
Thus, the polarization is parallel or antiparallel to the precession vector, and its magnitude is determined by the nonlinear gain saturation, conditioned by a finite number of atoms. In the present description, the self-organization gain compensates longitudinal relaxation in the stationary state, while transverse relaxation remains uncompensated.

The phenomenological term $\boldsymbol{\alpha}(\mathbf P)$ may, more generally, also contain transverse components leading to an additional rotation of the orientation vector. This possibility will be considered in Section~\ref{sec:bistability}.

\subsection{Reduced collective response and anisotropic narrowing}

The measured LD signal is the sum of the primary and secondary alignment responses,
\begin{equation}
S_A=S_{A,1}+S_{A,2},
\quad
S_{A,1}\propto Q_{1,xy},
\quad
S_{A,2}\propto Q_{2,xy},
\label{eq:signal_components}
\end{equation}
with the narrow resonances dominated by the collective $\mathsf{Q}_2$ contribution.

For the analytical description of the observed response, we introduce the complex tensorial quadrature,
\begin{equation}
q_j\equiv Q_{j,xy}-\mathrm{i}Q_{j,xz}
\quad
(j=1,2).
\label{eq:q_main}
\end{equation}

The slow collective mode can be represented either by its orientation
$\mathbf P$ or, in the rank-2 alignment sector, by the corresponding
reduced variable $q_2$. These two descriptions refer to the same
collective spontaneously polarized mode rather than to two independent
degrees of freedom. In the alignment representation, its dynamics is
written as

\begin{equation}
\dot q_2
=
-(\Gamma_P+\mathrm{i}\omega_x)q_2
+\kappa_{12}q_1.
\label{eq:q2_main}
\end{equation}
The variable $q_2$ should therefore be understood as a reduced
representation of the collective alignment response, rather than as an
additional independent dynamical mode.
Equation~\eqref{eq:q2_main} is not assumed to be an exact projection of the full $\mathsf Q_2$ dynamics; it is a phenomenological closure for the slow collective quadrature that captures the response of the collective alignment after faster degrees of freedom have been absorbed into effective parameters. In the stationary state,
\begin{equation}
Q_{2,xy}
=
\kappa_{12}
\frac{
\Gamma_PQ_{1,xy}-\omega_xQ_{1,xz}
}{
\Gamma_P^2+\omega_x^2
}.
\label{eq:Q2xy_main}
\end{equation}
Substituting the stationary primary-alignment components from Eqs.~\eqref{eq:Q1xy_main} and \eqref{eq:Q1xz_main} yields an expression for the observed LD signal:
\begin{equation}
\begin{aligned}
&S_{A,2}\propto{}
\kappa_{12}A_{10}\omega_x\omega_y\\
\hfill &\times\frac{
\left[
\Gamma_P
\left(
\Gamma_A^2+4\omega_x^2-2\omega_y^2
\right)
+
\Gamma_A
\left(
\Gamma_A^2+4\omega_x^2+\omega_y^2
\right)
\right]
}{
\left(
\Gamma_P^2+\omega_x^2
\right)
\left(
\Gamma_A^2+\omega_x^2+\omega_y^2
\right)
\left(
\Gamma_A^2+4\omega_x^2+4\omega_y^2
\right)
}.
\end{aligned}
\label{eq:SA2_main}
\end{equation}
The first factor in the denominator represents the slow collective response through the phenomenological rate $\Gamma_P$, while the remaining factors arise from the stationary dynamics of $\mathsf Q_1$. In the regime $\Gamma_P\ll\Gamma_A$, $|\omega_x|\ll\Gamma_A$, Eq.~\eqref{eq:SA2_main} reduces, up to a field-independent factor and relative corrections of order $\Gamma_P/\Gamma_A$ and $\omega_x^2/\Gamma_A^2$, to
\begin{equation}
S_{A,2}
\propto
\frac{
\omega_x
}{
\Gamma_P^2+\omega_x^2
}
\frac{
\omega_y
}{
(\Gamma_A/2)^2+\omega_y^2
}.
\label{eq:SA2_factorized_main}
\end{equation}
The corresponding resonance scale along $x$ is therefore governed by the slow collective rate, whereas the dependence on $\omega_y$ remains on the fast primary-alignment scale.

\begin{figure}[!t]
\includegraphics[width=1.0\columnwidth]{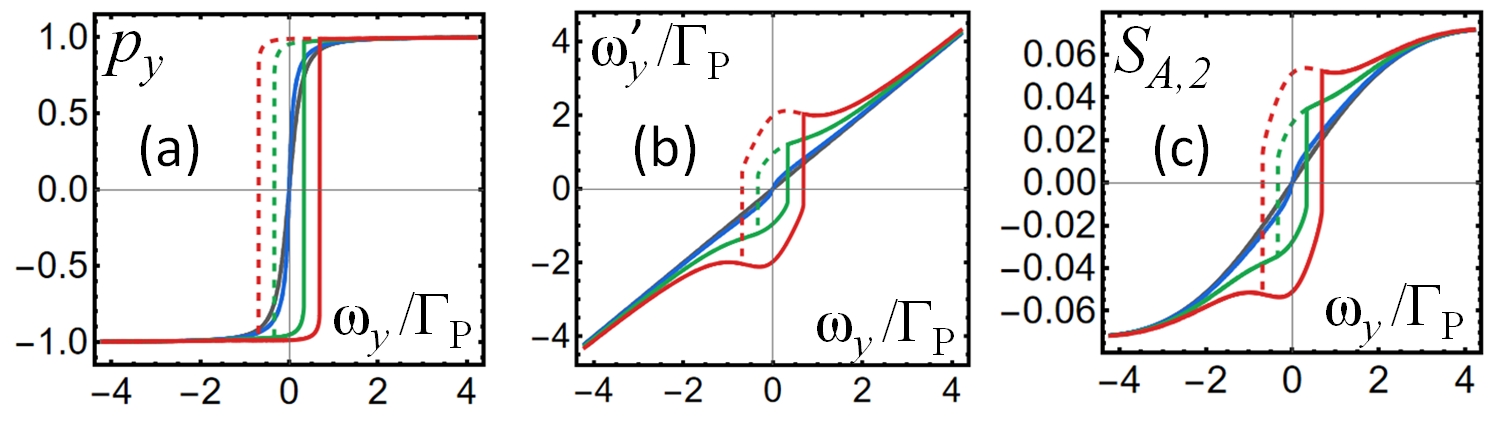}
\caption{
Quasistatic forward (solid lines) and reverse (dashed lines) scans of (a) the normalized polarization $p_y$, (b) the total effective field $\omega_y'/\Gamma_P$, and (c) the secondary-alignment signal $S_{A,2}$ (arb.\ units), as functions of the angular-frequency component $\omega_y/\Gamma_P$. The calculations use Eqs.~\eqref{eq:omega_y_eff_main},
\eqref{eq:main_Py_signed}, and~\eqref{eq:SA2_main},
with $\omega_y$ replaced by $\omega_y'$ in Eq.~\eqref{eq:SA2_main}, at $\omega_x/\Gamma_P=0.3$ and $\Gamma_A/\Gamma_P=10$. Gray, blue, green, and red curves correspond to $\beta/\Gamma_P=0$, $0.2$, $1.0$, and $2.0$, respectively.}
\label{fig:omega_y_prime}
\end{figure}

\begin{figure*}[!t]
\includegraphics[width=\linewidth]{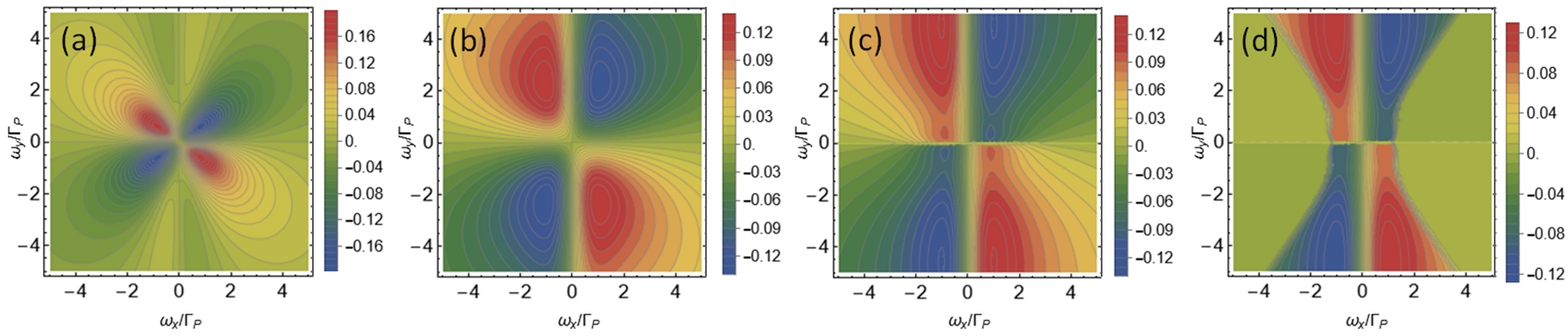}
\caption{Dependence of the $S_{A,2}$ signal (arb. units) on the normalized angular-frequency components $\omega_x/\Gamma_P$ and $\omega_y/\Gamma_P$ calculated using Eqs.~\eqref{eq:omega_y_eff_main},
\eqref{eq:main_Py_signed}, and~\eqref{eq:SA2_main}, with $\kappa_{12}A_{10}$ used as an arbitrary normalization. The panels correspond to (a) $\Gamma_A/\Gamma_P=1$, $\beta/\Gamma_P=0$; (b) $\Gamma_A/\Gamma_P=5$, $\beta/\Gamma_P=0$; and (c), (d) $\Gamma_A/\Gamma_P=10$, $\beta/\Gamma_P=2$. In panel (d), the condition $L_2(\cos\theta')\lesssim0$ is also included, with $\theta'$ denoting the angle between the total effective field $\mathbf B'$ and the $x$ axis.}
\label{fig:results}
\end{figure*}

\section{Effective field, bistability, and hysteresis}\label{sec:bistability}

The experimentally observed effective field, as well as the self-organization of polarization, can be related to the coupling between the primary and secondary alignments. For $\mathsf Q_1=\mathsf Q_{10}$, this coupling can be characterized by their scalar overlap Eq.~\eqref{eq:overlap}. Since the steady-state principal axis of $\mathsf Q_2$ is parallel to $\mathbf P$, in the stationary approximation $\theta$ is also the angle between $\mathbf P$ and the $x$ axis. The angular derivative of the overlap, used below to construct
the phenomenological torque, is
\begin{equation}
\frac{\partial\Delta Q}{\partial\theta}
=
-2A_{10}A_2\sin\theta\cos\theta.
\label{eq:overlap_derivative}
\end{equation}

Since the principal axis of $\mathsf Q_{10}$ coincides with the
$x$ axis, we introduce the unit vector $\mathbf n_\theta$
normal to the plane defined by $\mathbf e_x$ and $\mathbf n_P$
and use it to specify the direction of the phenomenological torque:
\begin{equation}
\mathbf n_\theta
=
\frac{\mathbf e_x\times\mathbf n_P}
{|\mathbf e_x\times\mathbf n_P|}
=
\frac{\mathbf e_x\times\mathbf n_P}{\sin\theta},
\end{equation}
and represent the resulting torque phenomenologically as
\begin{equation}
\boldsymbol{\tau}_{A}
=
-\kappa
\frac{\partial\Delta Q}{\partial\theta}\,
\mathbf n_\theta,
\label{eq:overlap_torque}
\end{equation}
where $\kappa$ is an effective coupling coefficient.

For $\mathbf P$ in the $xy$ plane and $P_y\neq0$,
\begin{equation}
\cos\theta=\frac{P_x}{P},
\quad
\sin\theta=\frac{|P_y|}{P},
\quad
\mathbf n_\theta=\operatorname{sgn}(P_y)\mathbf e_z.
\end{equation}

For the purpose of estimating the state-dependent torque, we use the idealized stretched-state relation
$A_2=\sigma P$ (Appendix~\ref{app:stretched_moments}), 
as an approximate closure for the collective secondary alignment.  
Using
$\cos\theta=P_x/P$,
$\sin\theta=|P_y|/P$,
and
$\mathbf n_\theta=\operatorname{sgn}(P_y)\mathbf e_z$,
we obtain
\begin{equation}
\boldsymbol{\tau}_{A}
=
2\kappa A_{10}A_2
\sin\theta\cos\theta\,
\mathbf n_\theta
=
2\kappa\sigma A_{10}
\frac{P_xP_y}{P}\,
\mathbf e_z.
\label{eq:overlap_torque_xy}
\end{equation}

Within the $xy$ plane, the same torque can be represented by an additional $y$-directed contribution to the precession vector, since
\begin{equation}
\boldsymbol{\tau}_{A}
=
\omega_{y,\mathrm{eff}}\,
\mathbf e_y\times\mathbf P
=
-\omega_{y,\mathrm{eff}}P_x\mathbf e_z.
\end{equation}
The term $\omega_{y,\mathrm{eff}}$ represents a state-dependent addition to the effective field: $\omega'_y=\omega_y+\omega_{y,\mathrm{eff}}$. Comparison with Eq.~\eqref{eq:overlap_torque_xy} gives
\begin{equation}
\omega_{y,\mathrm{eff}}
=
-2\kappa\sigma A_{10}
\frac{P_y}{P}.
\label{eq:omega_y_overlap}
\end{equation}

Because this contribution changes sign under $\mathbf P\rightarrow-\mathbf P$, the effective field $\boldsymbol{\omega}'$ and hence the collective dynamics are not invariant under this transformation. This state-dependent breaking of the $\mathbf P\rightarrow-\mathbf P$ symmetry provides an ingredient for the emergence of multiple self-consistent stationary states and, in particular, bistability.

For the stationary state $P=P_0$, this has the form
\begin{equation}
\omega_{y,\mathrm{eff}}
=
\beta p_y,
\qquad
p_y=\frac{P_y}{P_0},
\end{equation}
where the constant coupling coefficient can be written as
\begin{equation}
\beta
=
-2\kappa\sigma A_{10}>0.
\end{equation}
We therefore write the effective precession component as
\begin{equation}
\omega_y'
=
\omega_y+\omega_{y,\mathrm{eff}},
\qquad
\omega_{y,\mathrm{eff}}
=
\beta p_y.
\label{eq:omega_y_main}
\end{equation}

Since the slow collective response has a finite bandwidth characterized by $\Gamma_P$, the effective contribution is reduced when the applied $y$-directed precession frequency becomes comparable to $\Gamma_P$. We therefore use the self-consistent phenomenological form
\begin{equation}
\omega_{y,\mathrm{eff}}
=
\beta\frac{\Gamma_P^2}
{\Gamma_P^2+\omega_y^2}
\,p_y.
\label{eq:omega_y_eff_main}
\end{equation}

For $\mathbf P$ that adiabatically follows the effective field while preserving its magnitude, the two possible orientations can be written as
\begin{equation}
p_y
=
s\frac{\omega_y+\omega_{y,\mathrm{eff}}}
{\sqrt{\omega_x^2+(\omega_y+\omega_{y,\mathrm{eff}})^2}},
\quad s=\pm1.
\label{eq:main_Py_signed}
\end{equation}
Here $s$ distinguishes parallel and antiparallel following relative to the $y$-directed representation of the effective field; it is not the sign of the coupling coefficient. The response calculations shown below use the previously adopted $s=+1$ relation.

For sufficiently strong feedback, the self-consistency of the coupled equations produces hysteresis with abrupt switching between branches. 
The role of the angular replenishment condition in setting a threshold for memory effects is discussed in Sec.~\ref{sec:picture} and illustrated explicitly for constant effective-field shifts in Appendix~\ref{app:angular_threshold}.
The hysteresis width is not an independent parameter: in the limit $|\omega_x|\ll|\omega_y|$, it is controlled by $\beta$, which sets the characteristic effective-field scale [Fig.~\ref{fig:omega_y_prime}].

\section{Calculated response and comparison with experiment}\label{sec:results}

The dependences calculated from Eqs.~\eqref{eq:SA2_main}, \eqref{eq:omega_y_eff_main}, and~\eqref{eq:main_Py_signed} [Fig.~\ref{fig:results}] illustrate the two experimental regimes: panels (a) and (b) correspond to temperatures below the threshold $T_T$, whereas panels (c) and (d) correspond to $T>T_T$. At $\Gamma_A=\Gamma_P$ [Fig.~\ref{fig:results}(a)], the calculated signal exhibits a six-lobed structure similar to that of the standard alignment response \cite{Meraki_Elson_Ho_Akbar_Kozbial_Kolodynski_Jensen_2023}. With increasing $\Gamma_A/\Gamma_P$, the signal evolves toward a four-lobed structure, and the characteristic resonance scale along $y$ increases [Fig.~\ref{fig:results}(b)].

The narrowing along $x$ follows automatically from the temperature dependence of $\Gamma_P$ under strong spin exchange \cite{Happer_Tam_1977}.
In the system under consideration (see Section~\ref{sec:system}), an increase in temperature is accompanied by an increase in the spin-exchange relaxation rate $\Gamma_{\mathrm{SE}}$ in the high-field limit and, simultaneously, by a reduction of the effective spin-exchange relaxation rate $\Gamma_{\mathrm{SERF}}$ near zero field due to the SERF effect~\cite{Happer_Tam_1977, Seltzer_2008}. The temperature dependences of $\Gamma_{\mathrm{SE}}$ and $\Gamma_{\mathrm{SERF}}$ calculated according to Ref.~\cite{Seltzer_2008} are presented in Ref.~\cite{Petrenko_Vershovskii_2026b}. That work also shows that the experimentally measured resonance widths along the $y$ axis are governed by $\Gamma_{\mathrm{SE}}$, whereas those along the $x$ axis are governed by $\Gamma_{\mathrm{SERF}}$, resulting in the anomalous narrowing of the alignment resonances. A similar anisotropy is observed in Figs.~\ref{fig:results}(b) and~\ref{fig:results}(c). Accordingly, $\Gamma_A$ can be associated with $\Gamma_{\mathrm{SE}}$, while $\Gamma_P$ can be associated with $\Gamma_{\mathrm{SERF}}$. When analyzing Fig.~\ref{fig:results}, it should be kept in mind that the anisotropy of the ``classical'' alignment signal has the opposite sign [Fig.~\ref{fig:results}(a); see also Ref.~\cite{Meraki_Elson_Ho_Akbar_Kozbial_Kolodynski_Jensen_2023}].

In Fig.~\ref{fig:results}(d), we additionally account phenomenologically for the angular selectivity of the collective contribution by retaining it only in the range where $L_2(\cos\theta')\lesssim0$ (Eq.~\eqref{eq:legendre}), with $\theta'$ defined as the angle of the total effective field $\boldsymbol{\omega}'=(\omega_x, \omega_y', 0)$ relative to the $x$ axis.

These results reproduce static anomalous effects reported in \cite{Petrenko_Vershovskii_2026b}. The history-dependent response, including bistability and hysteresis, is reproduced by solving Eqs.~\eqref{eq:omega_y_eff_main}, \eqref{eq:main_Py_signed}, and~\eqref{eq:SA2_main} sequentially while retaining the preceding state of the system [Fig.~\ref{fig:omega_y_prime}].

The more complex system \cite{Petrenko_Vershovskii_2026a} can be studied using the same approach: a small addition of circularly polarized light creates a $P_z$ component which, under conditions where the vector $\mathbf P$ follows the effective field adiabatically, is equivalent to an effective $B_z$ component.

\section{Discussion}\label{sec:discussion}

The model developed above provides a common physical picture for the anomalous alignment response observed under strong spin exchange. The central element is a collective mode formed by spontaneous polarization and coupled to the optically generated primary alignment through a secondary alignment. This coupling transfers the slowly evolving collective dynamics to the alignment signal and, at the same time, allows the collective mode to retain alignment on a time scale much longer than the characteristic spin-exchange relaxation time. In this sense, the observed narrowing can be viewed as an effective preservation of the relevant alignment generated by the coupled dynamics, rather than as the suppression of spin-exchange relaxation of an independently evolving rank-2 moment.

A central feature of the model is that the different observed effects are interpreted in terms of the same tensor coupling, characterized by the scalar overlap $\mathsf Q_1:\mathsf Q_2$. In the uniaxial approximation, its angular dependence is determined by $L_2(\cos\theta)$ and provides a common geometric basis for replenishment of the collective mode and for the state-dependent torque. For $\mathsf Q_1=\mathsf Q_{10}$, the sign of the overlap selects the angular belt in which replenishment is assumed to sustain SP, while its angular derivative is used to construct the phenomenological torque responsible for the effective field. Thus, all anomalous effects are treated as related manifestations of alignment--collective-mode coupling. 

The anisotropic response follows naturally within the reduced description from the different dynamical roles of the primary and collective modes. The primary alignment responds on the fast scale $\Gamma_A^{-1}$, whereas the secondary alignment carries the slow collective scale $\Gamma_P^{-1}$. Their coupling produces a narrow response along $x$ while retaining the broader primary-alignment response along $y$. In the scale-separated regime, this distinction is explicit in Eq.~\eqref{eq:SA2_factorized_main}, where the characteristic frequency scales are $\Gamma_P$ and $\Gamma_A/2$, respectively.

The connection between orientation and secondary alignment is particularly simple for an ideal stretched-state component: its rank-1 and rank-2 moments have a common axis, and their amplitudes obey $A_2=\sigma P$ (Appendix~\ref{app:stretched_moments}). This relation provides a basis for associating the secondary alignment with the slowly relaxing collective polarization. It does not, however, require the instantaneous dynamics of the two moments to coincide in a general nonequilibrium state. Accordingly, the stretched-state relation is used here as an approximate closure when estimating the torque, whereas $q_2$ describes the reduced collective alignment response after other degrees of freedom have been absorbed into effective parameters.

The same overlap picture also provides an interpretation of bistability and hysteresis. The orientation of the spontaneous polarization determines the overlap with the primary alignment and hence the state-dependent torque. The associated effective field, together with relaxation and self-organization, determines the stationary polarization direction.  For sufficiently strong feedback, the resulting self-consistency permits more than one stationary solution for the same applied field and produces history-dependent switching under quasistatic scans, as illustrated in Fig.~\ref{fig:omega_y_prime}. The angular restriction additionally limits the region in which the collective contribution is retained [Fig.~\ref{fig:results}(d)].

The angular replenishment condition also provides a geometric explanation for the threshold behavior of memory effects. A sufficiently strong effective field makes the existence ranges of opposite polarized branches overlap, removing the interval of unavoidable SP suppression and allowing history-dependent branch selection, as illustrated in Appendix~\ref{app:angular_threshold}.

The quasistatic treatment describes the sequence of stationary responses rather than the detailed switching trajectory. In particular, near a vanishing total effective field or a branch termination, adiabatic following need not remain valid. Transient departures from the $xy$ plane, changes in the polarization magnitude, and fluctuations may then become relevant. The present calculation therefore does not determine switching times or the lifetime of a metastable branch. 

The present model is phenomenological: the gain sustaining SP, the transfer coefficient $\kappa_{12}$, and the bandwidth dependence of the effective-field term are not derived microscopically. Nevertheless, the model captures the principal features of the observed response within a common framework and connects them through a minimal number of physically motivated assumptions. Establishing the microscopic mechanism responsible for formation and replenishment of the collective mode, deriving the coupling between its moments, and extending the description to three-dimensional field geometries remain tasks for future work.

\section{Conclusions}\label{sec:conclusion}

We have developed a phenomenological model that accounts for the anomalous effects observed in our earlier experiments. In this model, the reported behavior results from the coupled dynamics of collective and single-atom degrees of freedom. The collective mode is viewed as a reservoir that is replenished by the primary alignment and retains the corresponding slow collective component of the alignment under rapid spin exchange.

The coupling between the primary and secondary alignments is characterized by their tensor overlap $\mathsf Q_1:\mathsf Q_2$. This single coupling factor provides a common basis for the observed effective conservation, anisotropic narrowing, and state-dependent effective field leading to bistability and hysteresis. Its angular dependence determines the range in which the collective contribution is effective, while its angular derivative produces the torque that feeds back on the collective polarization. The observed effects can therefore be understood as different manifestations of the same overlap-mediated dynamics.
The interplay between the effective field and the angular condition for replenishment further provides a geometric basis for the threshold onset of memory effects.

The observable therefore appears protected without any corresponding microscopic conservation law, amounting to an effective conservation law generated by the dynamics itself.

\section*{Acknowledgments}

The author gratefully acknowledges E.B.~Alexandrov, A.S.~Pazgalev, I.M.~Sokolov, and M.V.~Petrenko for fruitful discussions. AI-based tools were used to assist symbolic-algebra checks; the author is responsible for the final content. This research was funded by the baseline project FFUG-2024-0039 at the Ioffe Institute.

\appendix

\section{Orientation and alignment in a fully stretched state}
\label{app:stretched_moments}

The relation between the orientation and secondary alignment can be obtained directly from the moments of an ideal fully stretched state. We use dimensionless angular-momentum operators, with $\hbar$ absorbed into their definition, and consider
\begin{equation}
\rho_{\mathrm{st}}
=
|F,F\rangle_{\mathbf n}\,{}_{\mathbf n}\langle F,F|,
\end{equation}
where $\mathbf n$ is the polarization direction. Define the dipole moment and the symmetric traceless quadrupole tensor by
\begin{align}
M_i
&=
\langle F_i\rangle,
\\
T_{ij}
&=
\frac12\langle F_iF_j+F_jF_i\rangle
-\frac{F(F+1)}{3}\delta_{ij}.
\label{eq:app_stretched_definitions}
\end{align}

Choosing the quantization axis along $\mathbf n$, the stretched state satisfies
\begin{equation}
\langle F_n\rangle=F,
\qquad
\langle F_n^2\rangle=F^2.
\end{equation}
Axial symmetry and $\langle\mathbf F^2\rangle=F(F+1)$ give equal second moments along the two perpendicular axes,
\begin{equation}
\langle F_{\perp 1}^2\rangle
=
\langle F_{\perp 2}^2\rangle
=
\frac{F}{2}.
\end{equation}
The off-diagonal symmetrized moments vanish in this frame. Consequently, in arbitrary coordinates,
\begin{equation}
\frac12\langle F_iF_j+F_jF_i\rangle
=
\frac{F}{2}\delta_{ij}
+
F\left(F-\frac12\right)n_i n_j.
\end{equation}
Subtracting the isotropic part yields
\begin{align}
\mathbf M
&=
F\mathbf n,
\\
\mathsf T
&=
\frac{F(2F-1)}{2}
\left(\mathbf n\mathbf n^T-\frac{\mathsf I}{3}\right).
\label{eq:app_stretched_tensor}
\end{align}
Thus, the quadrupole axis coincides with the orientation direction, and its amplitude is positive for $F>1/2$.

To account for a variable weight of the collective component, let its contribution to the density operator be $f\rho_{\mathrm{st}}$, with $0\leq f\leq1$. This component need not constitute the entire atomic state: the primary alignment and other background contributions are treated separately. Linearity of expectation values gives
\begin{align}
\mathbf M_{\mathrm{c}}
&=
fF\mathbf n,
\\
\mathsf T_{\mathrm{c}}
&=
f\frac{F(2F-1)}{2}
\left(\mathbf n\mathbf n^T-\frac{\mathsf I}{3}\right).
\end{align}
Using the normalized orientation $\mathbf P=\mathbf M_{\mathrm{c}}/F$, we have $P=f$. If the secondary alignment is defined by $\mathsf Q_2=c_2\mathsf T_{\mathrm{c}}$, where $c_2>0$ specifies its normalization, then
\begin{equation}
\mathsf Q_2
=
A_2\left(\mathbf n\mathbf n^T-\frac{\mathsf I}{3}\right),
\qquad
A_2=\sigma P,
\label{eq:app_stretched_A2P}
\end{equation}
with
\begin{equation}
\sigma =c_2\frac{F(2F-1)}{2}.
\label{eq:app_stretched_a2}
\end{equation}

For the $F=4$ manifold of Cs, the stretched-state moments are
\begin{equation}
\mathbf M=4\mathbf n,
\qquad
\mathsf T
=
14\left(\mathbf n\mathbf n^T-\frac{\mathsf I}{3}\right),
\end{equation}
so that $\sigma=14c_2$. In a normalization where the quadrupole amplitude of a fully stretched state is unity, $c_2=1/14$ and $\sigma=1$.

Eq.~\eqref{eq:app_stretched_A2P} applies to a single ideal stretched-state component of fixed internal structure. It does not generally hold instantaneously for an arbitrary nonequilibrium collective state.

\section{Tensor  rotation}\label{app:tensor}
For a uniaxial tensor,
\begin{equation}
\mathsf Q_j=A_j(\mathbf n_j\mathbf n_j^T-\mathsf I/3).
\label{eq:SM_uniaxial}
\end{equation}
General dynamical tensors need not remain uniaxial. The rotation generator obeys $\Omega\mathbf v=\boldsymbol\omega\times\mathbf v$ and $\Omega^T=-\Omega$. An infinitesimal rotation $R=\mathsf I+\Omega\,dt+O(dt^2)$ gives
\begin{align}
\mathsf Q+d\mathsf Q
&=(\mathsf I+\Omega\,dt)\mathsf Q(\mathsf I+\Omega^Tdt)+O(dt^2)\nonumber\\
&=\mathsf Q+(\Omega\mathsf Q+\mathsf Q\Omega^T)dt+O(dt^2),
\end{align}
and consequently
\begin{equation}
\left.\dot{\mathsf Q}\right|_{\mathrm{rot}}=\Omega\mathsf Q-\mathsf Q\Omega.
\label{eq:SM_Q_rotation}
\end{equation}
The off-diagonal components are
\begin{align}
(\Omega\mathsf Q-\mathsf Q\Omega)_{xy}&=\omega_yQ_{yz}-\omega_xQ_{xz},\\
(\Omega\mathsf Q-\mathsf Q\Omega)_{xz}&=\omega_xQ_{xy}+\omega_y(Q_{zz}-Q_{xx}),\\
(\Omega\mathsf Q-\mathsf Q\Omega)_{yz}&=\omega_x(Q_{yy}-Q_{zz})-\omega_yQ_{xy}.
\label{eq:SM_rot_yz}
\end{align}

\section{Stationary primary alignment}\label{app:primary}
The pump-induced tensor and its dynamics are
\begin{align}
\mathsf Q_{10}&=A_{10}(\mathbf e_x\mathbf e_x^T-\mathsf I/3),\label{eq:SM_Qpump}\\
\dot{\mathsf Q}_1&=\Omega\mathsf Q_1-\mathsf Q_1\Omega-\Gamma_A(\mathsf Q_1-\mathsf Q_{10}),\label{eq:SM_Q1}
\end{align}
where $A_{10}<0$. Writing $Q_{ab}=Q_{1,ab}$, stationarity gives
\begin{align}
\Gamma_AQ_{xx}-2\omega_yQ_{xz}&=\tfrac23\Gamma_AA_{10},\label{eq:SM_Q1xx_eq}\\
\Gamma_AQ_{yy}+2\omega_xQ_{yz}&=-\tfrac13\Gamma_AA_{10},\label{eq:SM_Q1yy_eq}\\
\Gamma_AQ_{xy}-\omega_yQ_{yz}+\omega_xQ_{xz}&=0,\label{eq:SM_Q1xy_eq}\\
\Gamma_AQ_{xz}-\omega_xQ_{xy}-\omega_y(Q_{zz}-Q_{xx})&=0,\label{eq:SM_Q1xz_eq}\\
\Gamma_AQ_{yz}-\omega_x(Q_{yy}-Q_{zz})+\omega_yQ_{xy}&=0,\label{eq:SM_Q1yz_eq}
\end{align}
with $Q_{zz}=-Q_{xx}-Q_{yy}$. The rank-2 rotation generator has eigenvalues $im\omega$, $m=0,\pm1,\pm2$, where $\omega^2=\omega_x^2+\omega_y^2$. Define
\begin{equation}
D_A=(\Gamma_A^2+\omega_x^2+\omega_y^2)(\Gamma_A^2+4\omega_x^2+4\omega_y^2).
\label{eq:SM_DA}
\end{equation}
The solution is
\begin{align}
Q_{1,xy}&=A_{10}\frac{\omega_x\omega_y(\Gamma_A^2+4\omega_x^2-2\omega_y^2)}{D_A},\label{eq:SM_Q1xy}\\
Q_{1,xz}&=-A_{10}\frac{\Gamma_A\omega_y(\Gamma_A^2+4\omega_x^2+\omega_y^2)}{D_A},\label{eq:SM_Q1xz}\\
Q_{1,yz}&=-3A_{10}\frac{\Gamma_A\omega_x\omega_y^2}{D_A}.\label{eq:SM_Q1yz}
\end{align}
The diagonal components are
\begin{align}
Q_{1,xx}&=\frac{2A_{10}}{3D_A}\Bigl(\Gamma_A^4+5\Gamma_A^2\omega_x^2+2\Gamma_A^2\omega_y^2\nonumber\\
&\quad+4\omega_x^4-4\omega_x^2\omega_y^2+\omega_y^4\Bigr),\label{eq:SM_Q1xx}\\
Q_{1,yy}&=-\frac{A_{10}}{3D_A}\Bigl(\Gamma_A^4+5\Gamma_A^2\omega_x^2+5\Gamma_A^2\omega_y^2\nonumber\\
&\quad+4\omega_x^4-10\omega_x^2\omega_y^2+4\omega_y^4\Bigr),\label{eq:SM_Q1yy}
\end{align}
with $Q_{1,zz}=-Q_{1,xx}-Q_{1,yy}$. To first order in $\omega_y$, retaining the full $\omega_x$ dependence,
\begin{align}
Q_{1,xy}&=A_{10}\frac{\omega_x\omega_y}{\Gamma_A^2+\omega_x^2}+O(\omega_y^3),\label{eq:SM_Q1xy_weak}\\
Q_{1,xz}&=-A_{10}\frac{\Gamma_A\omega_y}{\Gamma_A^2+\omega_x^2}+O(\omega_y^3).\label{eq:SM_Q1xz_weak}
\end{align}

\section{Reduced collective response}\label{app:response}
Define
\begin{equation}
q_j=Q_{j,xy}-\mathrm{i}Q_{j,xz},
\quad j=1,2.
\end{equation}
The weak-$\omega_y$ primary response is
\begin{equation}
q_1=\frac{iA_{10}\omega_y}{\Gamma_A+i\omega_x}+O(\omega_y^3).\label{eq:SM_q1_weak}
\end{equation}
We assume the reduced equation
\begin{equation}
\dot q_2=-(\Gamma_P+i\omega_x)q_2+\kappa_{12}q_1.\label{eq:SM_q2_reduced}
\end{equation}
Here $\kappa_{12}$ is a real phenomenological transfer coefficient. This closure is not an exact projection of full tensor dynamics: only $Q_{2,xy}$ and $Q_{2,xz}$ are retained explicitly. Other tensor components and collective higher-rank moments enter through $\Gamma_P$ and $\kappa_{12}$. The collective contribution is assumed to disappear continuously:
\begin{equation}
\kappa_{12}\to0\quad\text{as}\quad P_0\to0.\label{eq:SM_kappa_limit}
\end{equation}
At stationarity,
\begin{equation}
q_2=\frac{\kappa_{12}}{\Gamma_P+i\omega_x}q_1,\label{eq:SM_q2_stationary}
\end{equation}
which gives
\begin{equation}
\begin{aligned}
q_2&=\frac{\kappa_{12}}{\Gamma_P^2+\omega_x^2}\Bigl[\Gamma_PQ_{1,xy}-\omega_xQ_{1,xz}\\
&\quad-i(\Gamma_PQ_{1,xz}+\omega_xQ_{1,xy})\Bigr].
\end{aligned}\label{eq:SM_q2_components}
\end{equation}
Taking the real part,
\begin{equation}
Q_{2,xy}=\kappa_{12}\frac{\Gamma_PQ_{1,xy}-\omega_xQ_{1,xz}}{\Gamma_P^2+\omega_x^2}.\label{eq:SM_Q2xy_reduced}
\end{equation}
Using the full primary solution, without expanding in $\omega_y$, define
\begin{align}
N_A&=\Gamma_P(\Gamma_A^2+4\omega_x^2-2\omega_y^2)\nonumber\\
&\quad+\Gamma_A(\Gamma_A^2+4\omega_x^2+\omega_y^2).\label{eq:SM_NA}
\end{align}
Then
\begin{align}
\Gamma_PQ_{1,xy}-\omega_xQ_{1,xz}&=\frac{A_{10}\omega_x\omega_yN_A}{D_A},\label{eq:SM_primary_combination}\\
Q_{2,xy}&=\frac{\kappa_{12}A_{10}\omega_x\omega_yN_A}{(\Gamma_P^2+\omega_x^2)D_A}.\label{eq:SM_Q2xy_exact}
\end{align}
Since $S_{A,2}\propto Q_{2,xy}$, this is the full signal in Eq.~\eqref{eq:SA2_main}, written using $D_A$ and $N_A$. It is the exact stationary result within the assumed closure, driven by the full primary response; it does not use Eq.~\eqref{eq:SM_q1_weak}. The slow denominator comes from the collective transfer factor, and the other factors from primary-alignment dynamics. To separate scales, take
\begin{equation}
\Gamma_P\ll\Gamma_A,\quad |\omega_x|\ll\Gamma_A,\label{eq:SM_scale_separation}
\end{equation}
without an expansion in $\omega_y/\Gamma_A$. Define $\epsilon_P=\Gamma_P/\Gamma_A$ and $\epsilon_x=\omega_x^2/\Gamma_A^2$. Exactly,
\begin{equation}
\begin{aligned}
N_A&=\Gamma_A(\Gamma_A^2+\omega_y^2)\\
&\quad+\Gamma_P(\Gamma_A^2-2\omega_y^2)
+4(\Gamma_A+\Gamma_P)\omega_x^2.
\end{aligned}
\end{equation}
Hence
\begin{equation}
N_A=\Gamma_A(\Gamma_A^2+\omega_y^2)[1+O(\epsilon_P)+O(\epsilon_x)],\label{eq:SM_NA_asymptotic}
\end{equation}
while
\begin{equation}
\begin{aligned}
D_A&=(\Gamma_A^2+\omega_y^2)(\Gamma_A^2+4\omega_y^2)\\
&\quad\times[1+O(\epsilon_x)].
\end{aligned}\label{eq:SM_DA_asymptotic}
\end{equation}
Cancellation gives
\begin{equation}
\begin{aligned}
Q_{2,xy}&=\frac{\kappa_{12}A_{10}\Gamma_A}{4}
\frac{\omega_x}{\Gamma_P^2+\omega_x^2}
\frac{\omega_y}{(\Gamma_A/2)^2+\omega_y^2}\\
&\quad\times[1+O(\epsilon_P)+O(\epsilon_x)].
\end{aligned}\label{eq:SM_Q2xy_factorized}
\end{equation}
Thus the signal is proportional, to leading order, to
\begin{equation}
S_{A,2}\propto\frac{\omega_x}{\Gamma_P^2+\omega_x^2}\frac{\omega_y}{(\Gamma_A/2)^2+\omega_y^2}.\label{eq:SM_central_signal}
\end{equation}

The slow scale is $\Gamma_P$ along $x$, and the primary scale $\Gamma_A/2$ remains along $y$. The total signal is
\begin{equation}
\begin{aligned}
S_A&=S_{A,1}+S_{A,2},\\
S_{A,1}&\propto Q_{1,xy},\quad S_{A,2}\propto Q_{2,xy}.
\end{aligned}\label{eq:SM_total_signal}
\end{equation}
As $P_0\to0$ and $\kappa_{12}\to0$, the conventional primary response remains for fixed positive relaxation rates.

\begin{figure}[!t]
\centering
\includegraphics[width=0.9\linewidth]{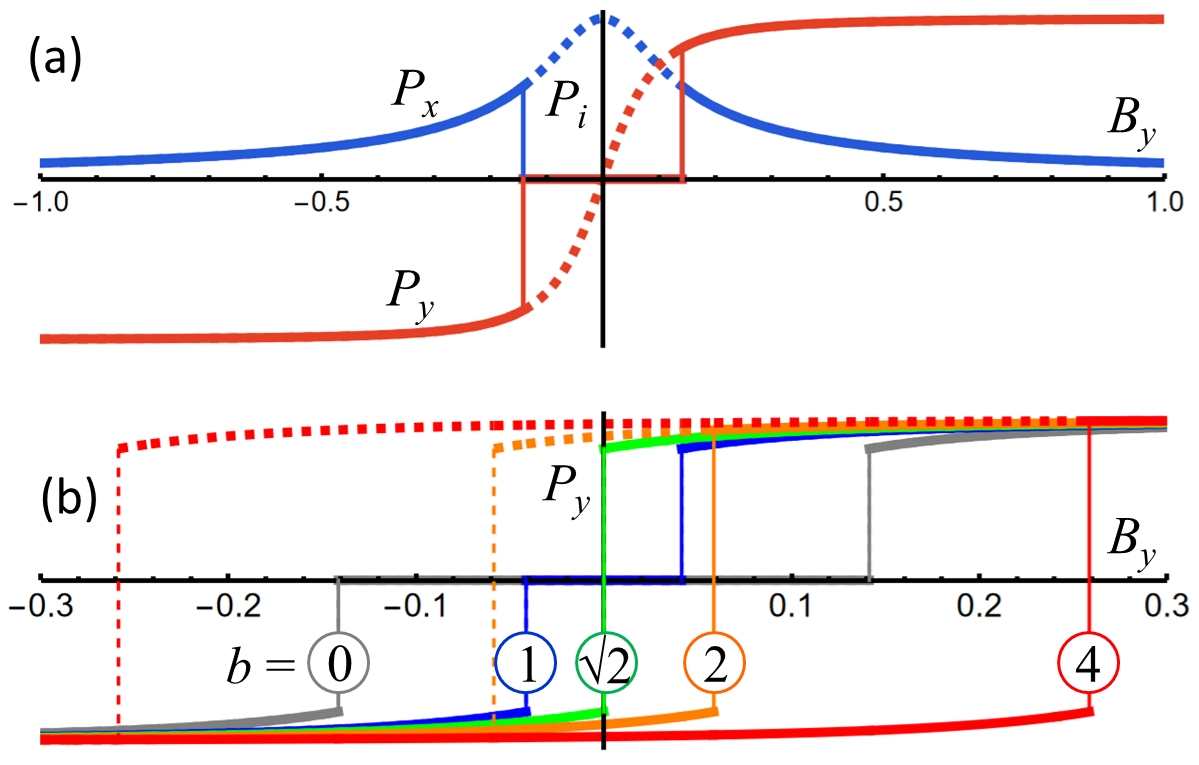}
\caption{Angular restriction and effective-field-induced branch overlap in a simplified quasistatic construction. The field components are expressed in common arbitrary units, with $B_x=0.1$, $B_z=0$, and $P_0=1$. (a) Components $P_x$ (blue) and $P_y$ (red) as functions of $B_y$ in the absence of an effective field. Dashed curves show Eq.~\eqref{eq:app_unrestricted_components}; solid curves include the angular restriction in Eq.~\eqref{eq:app_restricted_components}. Both components are set to zero for $|B_y|\leq\sqrt{2}|B_x|$. (b) Component $P_y$ under forward (solid) and reverse (dashed) scans for opposite effective-field shifts of constant magnitude $b|B_x|$. Gray, blue, green, orange and red curves correspond to $b=0,1,\sqrt{2},2,4$, respectively. A finite hysteresis loop appears for $b>\sqrt{2}$ under the assumed branch-continuation rule. 
}
\label{fig:angular_scans}
\end{figure}

\section{Angular restriction and the threshold for branch overlap}
\label{app:angular_threshold}

To illustrate the role of the angular restriction, we consider a quasistatic scan of $B_y$ at a fixed $B_x>0$, with $B_z=0$. The collective polarization has a constant magnitude $P_0$ wherever the replenishment condition is satisfied and is set to zero outside this region. This simplified construction describes the allowed polarized branches and their continuation during a scan, rather than the transient dynamics of their disappearance or switching.

\subsection{Response without an effective field}

In the absence of an effective field, we first choose the branch with $\mathbf P$ parallel to $\mathbf B$. Without the angular restriction, its components are
\begin{equation}
\begin{aligned}
P_x^{(0)}(B_y)&=P_0\frac{B_x}{\sqrt{B_x^2+B_y^2}},
\\
P_y^{(0)}(B_y)&=P_0\frac{B_y}{\sqrt{B_x^2+B_y^2}}.
\label{eq:app_unrestricted_components}
\end{aligned}
\end{equation}
The angle $\theta$ between $\mathbf P$ and the $x$ axis satisfies $\cos^2\theta=B_x^2/(B_x^2+B_y^2)$. The replenishment condition in Eq.~\eqref{eq:magic} therefore becomes
\begin{equation}
\cos^2\theta<\frac13
\quad\Longleftrightarrow\quad
|B_y|>\sqrt{2}|B_x|.
\label{eq:app_angular_field_condition}
\end{equation}
We adhere to the convention adopted in the model that, due to the positive feedback, the SP vector maintains a constant length throughout its entire range of existence—that is, within the angular range where its replenishment is positive.

Introducing the dimensionless constant $a=\sqrt{2}$ and the dimensionless variable $b_y=B_y/|B_x|$, we represent the angular restriction by
\begin{equation}
P_i(b_y)=
\begin{cases}
P_i^{(0)}(b_y), & |b_y|>a,\\
0, & |b_y|\leq a,
\end{cases}
\qquad i=x,y.
\label{eq:app_restricted_components}
\end{equation}
Figure~\ref{fig:angular_scans}(a) compares the restricted and unrestricted responses. Although the unrestricted vector rotates continuously through the $x$ direction, SP is not sustained in the interval $|b_y|\leq a$. In the present quasistatic construction, the polarization vanishes in this interval, and memory carried by it is lost. The plotted response after this interval again uses the parallel branch; the angular condition alone does not determine which polarization direction is selected when SP reappears.

\subsection{Response with opposite effective-field shifts}

We now introduce dimentionless coefficient $b$ characterizing two effective-field shifts of equal constant magnitude $b|B_x|\geq0$. The negative-polarization branch experiences the shift $-b|B_x|$, whereas the positive-polarization branch experiences the shift $+b|B_x|$:
\begin{equation}
b'_{y,-}=(b_y-b),
\qquad
b'_{y,+}=(b_y+b).
\label{eq:app_shifted_fields}
\end{equation}
Both branches considered below have $\mathbf P$ parallel to their respective total effective fields. This construction isolates the geometric effect of the field shift and does not use the full state-dependent expression in Eq.~\eqref{eq:omega_y_eff_main}.
The angular condition must be applied to the total effective field:
\begin{equation}
|b_y'|>a.
\label{eq:app_effective_angular_condition}
\end{equation}
The negative and positive branches are therefore
\begin{align}
P_y^-(b_y)&=P_0\frac{b_y-b}{\sqrt{1+(b_y-b)^2}},
& b_y&<b-a,
\label{eq:app_negative_branch}\\
P_y^+(b_y)&=P_0\frac{b_y+b}{\sqrt{1+(b_y+b)^2}},
& b_y&>a-b.
\label{eq:app_positive_branch}
\end{align}
For $b<a$, an interval remains in which neither of these branches is allowed:
\begin{equation}
b-a\leq b_y\leq a-b.
\label{eq:app_unpolarized_interval}
\end{equation}
Its width is $2(a-b)|B_x|$ and decreases with increasing effective field. At $b=a$, this interval contracts to a single point. For $b>a$, the existence ranges overlap:
\begin{equation}
a-b<b_y<b-a.
\label{eq:app_branch_overlap}
\end{equation}
Thus, two allowed polarized branches exist at the same applied field when
\begin{equation}
b>\sqrt{2}.
\label{eq:app_geometric_threshold}
\end{equation}
In particular, both branches are allowed at $B_y=0$. The effective field therefore removes the finite interval in which loss of SP is unavoidable in this construction.

To obtain the scan curves, we retain the selected branch until it reaches its angular boundary. In a forward scan, the negative branch is retained up to $B_y=b-a$; in a reverse scan, the positive branch is retained down to $B_y=a-b$. Above the threshold, the opposite branch is already allowed at each switching point, and the resulting hysteresis width is
\begin{equation}
\Delta b_y=2(b-a)=2\left(b-\sqrt{2}\right),
\qquad b>a.
\label{eq:app_hysteresis_width}
\end{equation}
Below the threshold, the polarization is set to zero between disappearance of one branch and appearance of the other, and the forward and reverse curves coincide. At the threshold, the hysteresis width remains zero.

Figure~\ref{fig:angular_scans}(b) shows these responses for $b=0,1,\sqrt{2},2,4$. The appearance of a finite loop above $b=\sqrt{2}$ illustrates the geometric threshold associated with the angular restriction. The vertical segments mark discontinuous changes between the plotted branches and do not describe intermediate stationary states.

This construction provides a simple basis for relating the onset of memory effects to overlap of the allowed polarized branches. It assumes that the selected branch persists until its angular boundary is reached. The dynamics and stability of the branches, as well as the polarization trajectory during switching, require a dynamical treatment. The threshold in Eq.~\eqref{eq:app_geometric_threshold} applies to the constant-shift construction used here; the corresponding threshold for the full model must be determined from its state-dependent effective field.

\bibliography{bibl}
\end{document}